\documentclass[conference,a4paper]{IEEEtran}
\usepackage{tikz}
\usepackage{pgfplots}
\usepackage{pgfplotstable}
\usepgfplotslibrary{fillbetween, patchplots, polar}
\tikzset{>=latex}
\usetikzlibrary{
  3d,
  arrows,
  arrows.meta,
  calc,
  matrix,
  plotmarks,
  positioning,
  shadings,
  shapes.geometric,
  decorations.pathmorphing
}
\tikzset{
  currentarrow/.style={
    -{Stealth[length=2.2mm,width=0.8mm]},
    line width=0.6pt,
    line cap=round
  }
}

\pgfplotsset{compat=newest}
\usepackage{standalone}    % 
\newlength\figurewidth
\newlength\figureheight
\usepackage{amsmath,amssymb,amsfonts}
\usepackage{graphicx} 
\usepackage{cite}
\usepackage{xcolor}
\providecommand*{\M}[1]
{\mathbf#1}        % matrices
\providecommand*{\V}[1]{\boldsymbol#1} % (physical) vectors
\providecommand*{\T}[1]{\mathrm{#1}}  % text 
\providecommand*{\eu}{\ensuremath{\T{e}}}  % e
\providecommand*{\ju}{\ensuremath{\T{j}}} % j
\renewcommand{\Re}{\operatorname{Re}} % The LaTeX standard is not ISO!
\providecommand*{\diff}{\operatorname{d}\!}  % d in integrals

\usepackage{xcolor}
\usepackage{balance}

\ifCLASSOPTIONcompsoc
 \usepackage[caption=false,font=normalsize,labelfont=sf,textfont=sf]{subfig}
\else
 \usepackage[caption=false,font=footnotesize]{subfig}
\fi

\begin{document}

\title{Focusing Efficiency for Large Intelligent Surfaces}

\author{
\IEEEauthorblockN{
Jiawang Li\textsuperscript{*},
Alireza Saberkari\textsuperscript{\(\dagger\)},
Buon Kiong Lau\textsuperscript{*},
Mats Gustafsson\textsuperscript{*}
}
\IEEEauthorblockA{
\textsuperscript{*}Dept. of Electrical and Information Technology, Lund University, 221 00 Lund, Sweden\\
\textsuperscript{\(\dagger\)}Dept. of Electrical Engineering, Linköping
University, 58183 Linköping, Sweden\\
jiawang.li@eit.lth.se
}
}

\maketitle

\begin{abstract}
This paper studies wireless power transmission (WPT) with large intelligent surfaces (LISs) in a single-user scenario, with emphasis on power-transfer efficiency. The optimal current distribution on the LIS is derived under a total-power constraint for an arbitrary receiving antenna, providing a theoretical characterization of the maximum achievable focusing efficiency. The analysis shows that the efficiency is strongly affected by the receiving radiation pattern, antenna orientation, and transmission distance. Clear performance differences are observed among practical antenna types for specific orientations, whereas directional averaging makes their average focusing efficiencies much closer. The theoretical findings are further validated by full-wave FEKO simulations, showing good agreement with the predictions of the proposed formulation. These results provide insight into the impact of receiver characteristics and propagation geometry on LIS-enabled WPT and offer a theoretical basis for efficient single-user power-transfer design.

\end{abstract}

\vskip0.5\baselineskip
\begin{IEEEkeywords}
large intelligent surface (LIS), wireless power transmission (WPT), near-field focusing, current optimization.
\end{IEEEkeywords}

\section{Introduction}
Large intelligent surfaces (LISs) have emerged as a promising technology for sixth-generation (6G) wireless systems in sub-10 GHz bands \cite{hu2018beyond1,hu2023design}. One important application is wireless power transfer (WPT), which can reduce battery dependence for low-power Internet of Things (IoT) devices. In large-scale LIS deployments, many devices are located in the near field \cite{selvan2017fraunhofer}, where focusing can provide substantially higher transfer efficiency than conventional far-field WPT \cite{kojima2020array,shen2020wireless}.

Considerable research has investigated near-field WPT \cite{borgiotti2003maximum,razavi2014maximum,kim2019efficiency,li2011design,chen2017maximum,lee2024design,mayer2024optimal,deutschmann2026physically}. Early studies derived optimal aperture excitations and efficiency bounds for planar transmitting and receiving structures \cite{borgiotti2003maximum,razavi2014maximum,kim2019efficiency}. Subsequent simulations and experiments confirmed that optimized field shaping can significantly improve power-transfer efficiency, including a 30\% improvement in implantable WPT \cite{chen2017maximum} and 72\% beam collection efficiency with a 256-element prototype \cite{lee2024design}. Recent work has further explored antenna placement, power allocation, and near-field focusing with physically large apertures \cite{mayer2024optimal,deutschmann2026physically}.

This paper develops a framework for single-user LIS-enabled WPT. First, we derive the optimal LIS current distribution under a total-power constraint and characterize the achievable focusing efficiency for arbitrary receiving antenna patterns. Second, we investigate the effects of antenna pattern, orientation, and transmission distance, with the theoretical results validated by full-wave FEKO simulations. Third, the framework provides a general basis for efficient single-user LIS-enabled WPT design.

\section{Modelling and error formulation
}\label{sec:Modelling and error formulation}

We consider a near-field WPT system in which an electrically large LIS transfers power to a compact receiving device. Because the physical size of the receiver is much smaller than its separation from the LIS, the field incident on the device can be locally approximated as a far-field excitation, even though the receiver itself lies within the near-field region of the LIS. This separation of scales allows the receiver to be described through its radiation pattern, while the LIS is modeled as a continuous radiating aperture.

\begin{figure}[t]
  \centering
\includegraphics[width=\linewidth]{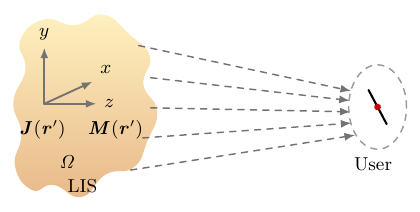}
  \caption{Schematic of the LIS transmitter and the receiving user. $\V J(\V r^\prime)$ and $\V M(\V r^\prime)$ denote the electric and magnetic current densities on the LIS surface. The center of LIS array is at the origin. The receiving antenna is shown as a half-wave dipole for illustration, but it can be an arbitrary receiving antenna element.}
  \label{fig:lis_receiver}
\end{figure}
As illustrated in Fig.~\ref{fig:lis_receiver}, the system consists of a single LIS transmitter with an arbitrary aperture geometry and one receiving antenna located in the service region. The shaded region \(\it \Omega\) represents the LIS aperture, over which equivalent electric surface current densities \(\V J(\V r^\prime)\) and/or magnetic surface current densities \(\V M(\V r^\prime)\), can in general be defined to generate the desired electromagnetic field at the receiver. In this work, however, only the electric surface current \(\V J(\V r^\prime)\) is considered. No specific antenna structure or radiation pattern is assumed at the receiving side. The half-wave dipole shown in the figure is used only for illustration, and the formulation applies generally to arbitrary receiving antennas characterized by their radiation patterns.

The WPT performance is evaluated using the focusing efficiency \(\eta_{\T{focus}}\), defined as the ratio of the RF power received by the antenna, prior to rectification, to the total power radiated by the LIS.

Solving the current-optimization problem for different levels of ohmic loss reveals how practical loss influences the optimized LIS excitation. At low loss levels, the optimal currents may exhibit rapidly varying and spatially irregular features. As the loss increases, these high-order variations are gradually suppressed, producing smoother current distributions that are more amenable to practical implementation \cite{gustafsson2019maximum}. The lossless or weakly lossy solution therefore serves as an upper-bound reference, whereas the lossy solutions provide a more realistic indication of the efficiency achievable under implementation constraints.

The optimized solution is also compared with the time-reversal (TR) excitation~\cite{fink2002time}. As the conductor loss increases, the optimal current distribution progressively approaches the TR solution, indicating that TR provides a simple and robust approximation in sufficiently lossy regimes. In contrast, under low-loss conditions, direct optimization can achieve higher transfer efficiency, although the resulting current distributions may be more difficult to realize physically.

Our goal is therefore to determine the maximum WPT focusing efficiency \(\eta_{\T{focus}}\) from an LIS transmitter to a user with a given radiation pattern under a total-power constraint. The problem can be formulated as
\begin{equation}\label{eq:local_tptal_power_constraints}
\begin{aligned}
 & {\T{max}} && \text{Receiving power at the user}\\
& \T{s.t.}\ && \text{Total power constraint.}
\end{aligned}
\end{equation}

According to the Poynting Theorem \cite{Harrington},
\begin{multline}\label{eq:PoyntingTheorem}
-\frac{1}{2}\int_{\it \Omega}\V{J}_{\T s}^{*}\cdot \V{E}_{\T s}\,\diff \T S'
=
\frac{1}{4}\oint_{\partial V}
\big(\V{E}\times\V{H}^{*}+\V{E}^{*}\times\V{H}\big)\cdot \hat{\M n}\,\diff \T S \\
+ \ju\,2\omega\big(\widetilde{W}_{\T m}-\widetilde{W}_{\T e}\big).
\end{multline}
where $\V J_\T s$ is the current density and $\V E_\T s$ is the tangential electric field of the transmission surface $\it \Omega$. $\diff \T S^\prime$ denotes the differential area, $\T S$ is the closed surface that surrounds the transmission region of volume $V$. $\V E$ and $\V H$ are the electric field and the magnetic field generated by the surface $\T S$. $\omega$ is the angular frequency.
$\widetilde{W}_\T m$ and $\widetilde{W}_\T e$ are the stored magnetic and electrical energy. For the sake of simplicity, electric current is taken here as an illustrative example.

Based on the MoM, the current density $\V{J}_\T s(\V{r^\prime})$ in $\it \Omega$ is expanded in a set of basis functions $\V \psi_n(\V r^\prime)$,
\begin{equation}\label{eq:MoM}
\V{J}_\T s(\V{r^\prime})
\approx
\sum_{n=1}^{N} I_n\,{\V \psi}_n(\V{r^\prime})
\end{equation}
 The current density $\V J_\T s(\V{r}')$ and electric field $\V E_\T s(\V{r}')$ is linear mapped \cite{eibert2015electromagnetic}, i.e. $
\V E_\T s(\V{r}') = \mathcal{L}\{\V J_\T s(\V{r}')\}
$, and
\begin{equation}
\mathcal{L}
=
-\ju\omega\mu
\iint_{\Omega}
\left(\M{I}_3+\frac{1}{k^{-2}}\nabla\nabla\right)
g(\V{r},\V{r^\prime})
\,\diff \T S' .
\end{equation}
where $\mu$
 is the permeability in vacuum and $g(\V{r},\V{r}')
=
{\eu^{-\ju k|\V{r}-\V{r}'|}}/({4\pi | \V{r}-\V{r}'|})$.
Substituting the expansion \eqref{eq:MoM} yields
\begin{equation}
\V{E}_\T s(\V{r^\prime})
=
\sum_{n=1}^{N} I_n(\V r_n^\prime)\,\mathcal{L}\{\V{\psi}_n(\V r^\prime)\}
\triangleq
\sum_{n=1}^{N} I_n\,\V{\Lambda}_n(\V{r^\prime}),
\end{equation}
where $\V{\Lambda}_n(\V{r^\prime})$ denotes the field produced when the $n$th basis function is excited alone.
Then the induced power $\it P$ can be expressed as
\begin{equation}
P
= \frac{1}{2}\T{Re}\left\{
\int_{\it \Omega}
\left(\sum_{m} I_m^{*}\,\V \psi_m^{*}(\V r^\prime)\right)\cdot
\left(\sum_{n} I_n\,\M \Lambda_n(\V r^\prime)\right)
\,\diff \T S^\prime
\right\}.
\end{equation}

We expand the product and interchange summation and integration, 
\begin{equation}\label{eq:total power}
P
= \frac{1}{2}\Re\left\{
\sum_{m}\sum_{n} I_m^{*}I_n
\int_{\it \Omega}
\V \psi_m^{*}(\V r^\prime)\cdot \M \Lambda_n(\V r^\prime)
\,\diff \T S^\prime
\right\}.
\end{equation}
Define
\begin{equation}
\M Z_{mn}
\triangleq
\int_{\it\Omega}\V \psi_m^{*}(\V r^\prime)\cdot \V \Lambda_n(\V r^\prime)\,\diff \T S^\prime,
\end{equation}
Then we can rewrite~(\ref{eq:total power}) as
\begin{equation}
P \approx \frac{1}{2}\T{Re}\left\{\M I^{     \T H}\M Z\,\M I\right\}.
\end{equation}

For any complex matrix $\M Z$ and vector $\M I$,
$
\T{Re}\left\{\M I^{\T H}\M Z\,\mathbf I\right\}
=
\M I^{\T H}\T{Re}\{\M Z\}\,\M I,
\T{Re}\{\M Z\}=\frac{1}{2}\left(\M Z+\M Z^{\T H}\right).
$ Therefore,
\begin{equation}
P \approx \frac{1}{2}\,\M{I}^{\T H}\M{R}\M{I}
\qquad \text{with}\qquad
\M{R}=\Re\{\M{Z}\}.
\end{equation}
$\M R$ should be positive definite matrix \cite{gustafsson2016antenna}. Next, we also express the transmission surface conductor loss $P_{\Omega}$ in the form of a function of basis functions as
\begin{equation}
P_{\Omega}
= \frac{1}{2}\int_{\it \Omega} R_\T s(\V{r^\prime})\,\left|\V{J}_\T s\right|^2\,\diff \T S^\prime ,
\label{eq:Ploss_surface}
\end{equation}
where $R_\T s(\V{r^\prime})$ represents the ohmic loss. Here, we assume that the ohmic loss is uniform across the entire transmission surface. We can use
$R_{\T s}=\sqrt{{\pi f \mu}/{\sigma}}
$ ($f$ is the working frequency and $\sigma$ is the conductivity)
to calculate it for common conductors such as copper, with a typical value of $0.01~\Omega/\square$. Similar to \eqref{eq:total power},
\begin{equation}
P_{\Omega}
= \frac{1}{2}\sum_m\sum_n I_m I_n^{*}
\underbrace{\int_{\it\Omega} R_\T s(\V{r^\prime})\,\V{\psi}_m(\V{r^\prime})\cdot\V{\psi}_n^{*}(\V{r^\prime})\,\diff \T S^\prime}_{\triangleq\,[\M{R}_{\Omega}]_{mn}} .
\label{eq:Ploss_matrix_elements}
\end{equation}
Power dissipation by the conductor can be expressed as with respect with current density
\begin{equation}
\M C \approx\frac{1}{2}\,\M{I}^{\T{H}}\M{R}_{\Omega}\M{I}.
\label{eq:Ploss_quadratic}
\end{equation}
Total power is
\begin{equation}
P_{\T{tot}}=\frac{1}{2}\,\M I^{\T H} \M R_\T {tot}\M I,
\qquad
\M R_\T {tot}=\M R+\M{R}_{\Omega} .
\end{equation}
For the receiving power from the receiver, we need to build a connection with the transmission surface. The receiving power $P_{\T r}$ can be expressed as
\begin{equation}
P_{\T r}={\left|V_{\T r}\right|^{2}}/(8R_\T {in})
={\left|V_{\T r}\right|^{2}}/(8R_\T L)
\end{equation}
where $R_\T {in}$ and $R_\T L$ are the matched input resistance and load resistance for the receiver. $V_{\T r}$ is the open circuit voltage at receiver, which is
\begin{equation}
V_{\T r}
=
-\frac{1}{I_\T r}
\int_{\T S_\T r}
\left(\V{E}_\T s \cdot \V{J}_\T r\right)\,\diff \T S
=-\frac{1}{I_\T r}
\int_{\it \Omega}
\left(\V{E}_\T r \cdot \V{J}_\T s\right)\,\diff \T S^\prime
\end{equation}
where $\T S_\T r$ is the surface enclose the receiving antenna and $I_\T r$ is the induced current when it it used as transmitting antenna. The reciprocity theorem is used to express the voltage as a function of the current density at the transmitting surface \cite{Harrington}.
Expand $V_{\T r}$ into basis functions
\begin{equation}
V_{\T r} = -\frac{1}{I_\T r}\sum_{n=1}^{N}\int_{\it{\Omega}}
\big( I_n\,\hat{\M{u}}_\T J \big)\cdot \V{E}_\T r\,\diff \T S^\prime ,
\end{equation}
where $\hat{\M{u}}_\T J$ is the unit vector parallel to the transmission surface. Then it can be expressed as
\begin{equation}
V_{\T r} \approx  -\frac{1}{I_\T r}\M{w}^{\T H}\M{I}
= -\frac{1}{I_\T r}\sum_{n=1}^{N}{w^*_n}\,I_n
= -\frac{1}{I_\T r}\sum_{n=1}^{N}\left(\hat{\M{u}}_\T J^{\T{T}}\V{E}_\T r\Delta \T S\right) I_n.
\end{equation}
Therefore, the $\eta_\T {focus}$ can be simplified as
\begin{equation}
\eta_\T {focus}
=
\frac{1}{8R_\T L\,|I_\T r|^{2}\,P}
\left|
\int_{\it \Omega} \V{E}_\T r \cdot \V{J}_\T s\, \diff \T S^\prime
\right|^{2}.
\end{equation}
 For a lossless receiving antenna, the input power equals the radiated power, i.e.,
$P^{\T r}_{\T{in}} = P^{\T r}_{\T{rad}} .
$
The input power can be written as
\begin{equation}
P^{\T r}_{\T{in}} = \frac{1}{2} R_\T {in} |I_\T r|^2
= \frac{1}{2} R_\T L |I_\T r|^2,
\end{equation}
and the radiated power is
\begin{equation}
P^{\T r}_{\T{rad}}
= \int_{4\pi} \frac{\left| {E_\T r}(\theta,\phi)\right|^{2}}{2\eta_{0}}\, r^{2}\,\diff \Omega_\T r ,
\end{equation}
where
$\left| {E_\T r}(\theta,\phi)\right|^{2}
= \left|E_{\theta}(\theta,\phi)\right|^{2}+\left|E_{\phi}(\theta,\phi)\right|^{2}$ and $\eta_0$ is the wave
impedance in the free space. where $\theta$ is the polar angle measured from the $+z$ axis ($0\le\theta\le\pi$), $\phi$ is the azimuth angle in the $xy$-plane about the $z$ axis ($0\le\phi<2\pi$), $\diff\Omega_r$ is the differential solid-angle element (typically $\diff\Omega_r=\sin\theta\,\diff\theta\,\diff\phi$), and $E(\theta,\phi)$ is the complex amplitude of the far-field electric field in the direction $(\theta,\phi)$. ${E_\T r}(\theta,\phi)$ can be obtained in through: 1) analytical solutions, such as half-wave dipole antenna \cite{balanis2016antenna}, and circular patch antenna \cite{long1978theory} or 2) full-wave simulations, using software such as Altair FEKO and CST Microwave Studio Suite.

Therefore, the efficiency can be simplified as
\begin{equation}
\eta_\T {focus}
=
\frac{\eta_0}{8P}
\frac{
\left|\displaystyle\int_{\it \Omega}\V{E}_\T r\cdot\V{J}_\T s\,\diff \T S^\prime \right|^2
}{
\displaystyle\int_{4\pi}\left(|E_\theta(\theta,\phi)|^2+|E_\phi(\theta,\phi)|^2\right) r^2\,\diff \Omega_\T r
}
\end{equation}

The problem can be formulated as maximizing the WPT efficiency under the total-power constraint, which leads to
\begin{equation}
\begin{aligned}
\max_{\M J}\quad & \eta_\T {focus} \\
\text{s.t.}\quad & \frac{1}{2}\,\M I^{\T H}\M{R}_{\text{tot}} \M I = P_\T{tot} .
\end{aligned}
\end{equation}
This setup is closely analogous to the maximum antenna gain problem \cite{gustafsson2019maximum}, which seeks to maximize radiation intensity in a given direction under a constraint on the total input power, including both radiated and ohmic power. It is important to note that under this formulation, LIS will radiate bidirectionally, causing the receiver to only receive moderate power. Therefore, the final calculated efficiency needs to be multiplied by 2. The entire solution process employs the method of moments (MoM) \cite{harrington1993field} and the Rao–Wilton–Glisson (RWG) triangular basis functions \cite{rao1982rwg}. To ensure accuracy, at least 10 elements are set within each wavelength.

\begin{figure}[t]
  \centering
\includegraphics[width=\linewidth]{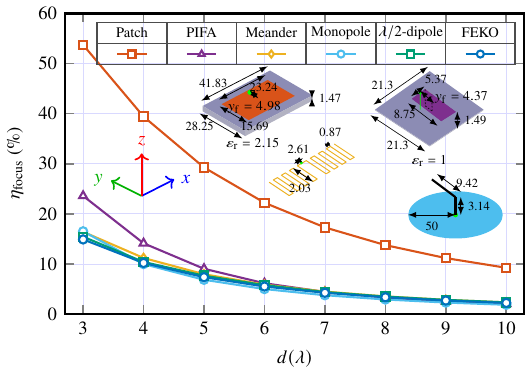}
  \caption{The maximum $\eta_\T {focus}$ achieved by different antenna types varies with distance. The LIS is placed at a distance $d$ from the receiving antenna along the $z$-direction. All antennas operate at 6~GHz and all parameters are in mm. The green dots indicate the feeding locations, and \(y_\T f\) denotes the offset of the feed position from the antenna center for both the patch antenna and the PIFA. In the PIFA, the dashed lines indicate the shorting patch, whose size is \(1.52 \times 1.52~\mathrm{mm}^2\). The monopole antenna adopts a folded L-shaped structure. A $\lambda/2$ dipole is evaluated, and the results are compared with those obtained from Altair FEKO 2023 MoM solver. In the FEKO simulation, the LIS is modeled as a planar array of discretely excited short dipoles aligned along $x$-direction, where each dipole has a length of $\lambda/20$ and the inter-element spacing is $0.4\lambda$ along both the $x$- and $y$-directions.}
  \label{fig:antenna_type}
\end{figure}
Figure~\ref{fig:antenna_type} shows the transmission efficiency as a function of distance with a LIS plane aperture $4\lambda \times 4\lambda$ for different receiving antenna types. It shows that the $\eta_\T {focus}$ decreases monotonically with the transmission distance \(d\) (along $z$-direction). The FEKO results for $\lambda/2$-dipole agree well with the theoretical curves, which confirms the effectiveness of LIS-enabled near-field WPT, especially in the short-range region (i.e., several wavelengths).

Moreover, the focusing efficiency is largely determined by the ratio between the side length of the square LIS and the focusing distance \(d\). For the \(\lambda/2\)-dipole, configurations with the same aperture-to-distance ratio yield approximately the same \(\eta_{\T{focus}}\). For example, \(\eta_{\T{focus}}\) is about \(3.5\%\) for a ratio of \(0.5\) (\(d=8\lambda\)), while it decreases to about \(1\%\) when the ratio is reduced to \(0.25\) (\(d=16\lambda\)). As this ratio becomes very large, the LIS approaches coverage of the half-space surrounding the dipole and can therefore collect up to half of its radiated power. This limiting behavior can also be interpreted through reciprocity~\cite{constantine2003antenna}: if the dipole acts as the transmitter and the LIS as the receiver, an LIS covering the entire half-space intercepts one half of the dipole radiation.

To demonstrate the applicability of the proposed framework to practical receiving antennas, Fig.~\ref{fig:antenna_type} compares \(\eta_{\T{focus}}\) for four representative antenna types: a probe-fed patch antenna, a probe-fed planar inverted-F antenna (PIFA), a meander dipole, and a folded L-shaped monopole. The electric fields over the LIS aperture at different separation distances can be obtained either directly from full-wave simulations or extrapolated from a single simulation using spherical-wave expansion (SWE) \cite{ludwig2003near}. In all cases, the antenna normal direction, i.e., the \(z\)-axis, is oriented toward the LIS.

Among the considered antennas, the patch antenna achieves the highest focusing efficiency throughout the investigated distance range. Although \(\eta_{\T{focus}}\) decreases with increasing distance for all antenna types, the patch retains a clear advantage, highlighting the importance of the receiver radiation pattern in near-field WPT. The PIFA, dipole, and monopole exhibit lower and relatively similar efficiencies, particularly at larger distances. These results illustrate how the proposed method can be used to evaluate the WPT performance of practical receiving devices under different antenna configurations and operating distances.
\begin{figure}[t]
  \centering
\includegraphics[width=\linewidth]{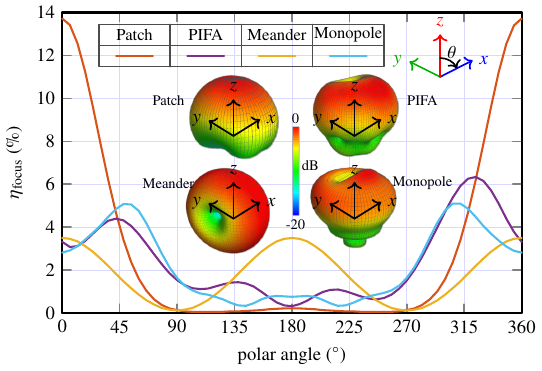}
  \caption{$\eta_\T {focus}$ variation of different antenna types (with different radiation patterns) under different polar angles at $d = 8\lambda$. The \(z\)-direction is normal to the antenna plane, and the polar angle \(\theta\) is measured from the \(z\)-axis toward the \(x\)-axis.}
  \label{fig:Orientation_antenna}
\end{figure}
Figure~\ref{fig:Orientation_antenna} shows the maximum achievable $\eta_\T {focus}$ of the device under different orientation angles at $d = 8\lambda$. The antenna parameters are the same as those in Fig.~\ref{fig:antenna_type}, and the variation of $\eta_\T {focus}$ is evaluated for different polar angles \(\theta\) (defined in Fig.~\ref{fig:Orientation_antenna}). The results show that the maximum $\eta_\T {focus}$ strongly depends on the device orientation, and the sensitivity to rotation varies with the antenna type. 

The patch antenna provides the highest peak \(\eta_{\T{focus}}\) at \(0^\circ\), but is also the most sensitive to orientation, with its efficiency falling below \(1\%\) at certain polar angles before recovering at larger angles. The PIFA follows a similar trend, showing high efficiency at small angles, a pronounced degradation at intermediate angles, and a subsequent recovery. The dipole exhibits a lower peak efficiency and remains relatively weak over the considered angular range, whereas the monopole shows a more moderate angular variation and comparatively stable performance.

These results demonstrate that receiver orientation can substantially affect LIS-enabled WPT performance. Practical receiver design should therefore consider not only the maximum achievable efficiency but also its sensitivity to orientation. The same framework can also be extended to curved transmitting surfaces, such as conformal arrays, although such configurations are not considered here.

To evaluate the LIS performance under arbitrary receiving-antenna orientations, the antenna radiation is represented by a spherical-wave expansion (SWE) in its body-fixed coordinate system. The expansion coefficients are extracted from the tangential near-field components $E_\theta$ and $E_\phi$, sampled on a source sphere of radius $r_{\mathrm{src}}$, using a regularized least-squares procedure. The radiated field is expressed as \cite{hansen1988spherical}
\begin{equation}
\mathbf{E}_0(\V{r})
=
\sum_{n=1}^{N_{\max}}\sum_{m=-n}^{n}
\left[
a_{nm}\mathbf{M}_{nm}^{(2)}(k\V{r})
+
b_{nm}\mathbf{N}_{nm}^{(2)}(k\V{r})
\right],
\end{equation}
where $a_{nm}$ and $b_{nm}$ are the expansion coefficients, $\mathbf{M}_{nm}^{(2)}$ and $\mathbf{N}_{nm}^{(2)}$ are the outgoing spherical vector wave functions, and $N_{\max}$ denotes the truncation order. Once these coefficients are determined, the radiated field can be reconstructed at arbitrary observation points without additional full-wave simulations.

For an arbitrarily rotated antenna, the field is obtained through a rigid-body coordinate transformation. Let $\M{Q}$ denote the rotation matrix from the antenna body coordinate system to the global coordinate system. The rotated field at an LIS observation point $\V{r}_{\T{LIS}}$ is then evaluated as
\begin{equation}
\M{E}_{\T{rot}}(\V{r}_{\T{LIS}})
=
\M{Q}\,
\M{E}_0\!\left(\M{Q}^{\T T}\V{r}_{\T{LIS}}\right),
\end{equation}
where $\M{Q}=\M{Q}_{\T y}(\theta_{\T{rot}})
\M{Q}_{\T z}(\phi_{\T{rot}})$ represents successive rotations about the $z$- and $y$-axes. This transformation allows the same SWE coefficients to be reused for different antenna orientations.

The LIS is modeled as a planar aperture located at $z=d$. At each sampling point $\V{r}_p=[x_p,y_p,d]^{\T T}$, the reconstructed field is converted into Cartesian components, from which the tangential components $E_\T x$ and $E_\T y$ are extracted. The incident power on the LIS is approximated by
\begin{equation}
P_{\T{inc,LIS}}
\approx
\sum_p
\frac{|E_\T x(\V{r}_p)|^2+
|E_\T y(\V{r}_p)|^2}{2\eta_0}
\Delta A_p,
\end{equation}
where $\Delta A_p$ is the surface element associated with the $p$th sampling point and $\eta_0$ is the free-space impedance.

To statistically assess the influence of antenna orientation, the orientation-averaged focusing efficiency $\eta_{\T{focus}}$ is evaluated using uniform sphere sampling (USS). The orientations are generated by uniformly sampling $\cos\theta_{\T{rot}}$ over $[-1,1]$ and $\phi_{\T{rot}}$ over $[0,2\pi]$. For each orientation, the incident field is reconstructed using the SWE, and the corresponding focusing efficiency is calculated. Despite their different angular radiation characteristics, all four antennas achieve an average focusing efficiency of approximately $2\%$. This indicates that the performance differences among the considered antennas become considerably less pronounced when their orientations are uniformly distributed over the sphere. The estimated averages are expected to converge further as the number of Monte Carlo samples increases.

This convergence result can be interpreted from the received power on the LIS. Specifically, the received power is determined by the normal component of the time-averaged Poynting vector integrated over the receiving surface. In the far field, using 
$\V{E}\times\V{H}^*\approx \hat{\V{r}}|\V{F}|^2/r^2$, it can be written as
\begin{equation}
P_{\T{rec}}=\frac{1}{2}\Re\int_{\T{LIS}}
(\V{E}\times \V{H}^*)\cdot \hat{\V{n}}\,\diff \T S
\approx\frac{1}{2}\int_{\T{LIS}}\frac{|\V{F}|^2}{r^2}
\left(\hat{\V{r}}\cdot\hat{\V{n}}\right)
\,\diff \T S .
\end{equation}
where $\V{F}$ denotes the far-field radiation vector, $r$ is the distance from the source to the observation point on the receiving surface, and $\hat{\V{r}}$ is the corresponding radial unit vector. The factor $\hat{\V{r}}\cdot\hat{\V{n}}$ represents the projection of the propagation direction onto the surface normal, and therefore accounts for the effective aperture seen by the incoming wave.

After averaging over antenna orientations, the pattern-dependent term is absorbed into the total radiated power, and the remaining geometric contribution is governed by
\begin{equation}
P_{\T{rec}}=\int_{\T{LIS}}\int_{4\pi}
\frac{|\V{F}(\V{r}')|^2}{4\pi r^2}\,
\hat{\V{r}}\cdot\hat{\M{n}}\,
\diff \Omega'\,\diff \T S
=
P_\T {rad}
\int_{\T{LIS}}
{\hat{\V{r}}\cdot\hat{\M{n}}}\,\diff \T S,
\end{equation}
where \(P_\T {rad}\) denotes the total radiated power. which is precisely the solid angle subtended by the receiving surface as seen from the source. Therefore, the convergence result indicates that, after directional averaging, the received power depends only on how large the LIS appears from the source location, rather than on the detailed radiation pattern itself. In the far field, this can also be approximated by \(\eta\approx A/(4\pi d^2)\) with \(A=16\lambda^2\), yielding \(\eta\approx 1/(16\pi)\approx 0.0199\), which is very close to the USS result.

\section{Conclusion}

This paper investigated LIS-enabled WPT with a focus on single-user power transfer. An optimization framework was established to obtain the LIS current distribution that maximizes the received power under a total-power constraint for an arbitrary receiving antenna pattern. Based on this formulation, the achievable power-transfer efficiency was characterized for different receiver configurations. Numerical results, together with full-wave FEKO simulations, confirmed the theoretical predictions and demonstrated the influence of receiver type, orientation, and transmission distance on the resulting focusing efficiency.


\bibliographystyle{IEEEtran}
\bibliography{citations}

\end{document}